\documentclass[12pt]{article}
\usepackage{graphicx}

\newcommand{\y}{Y(4260)}
\newcommand{\z}{Z_c(3900)}
\newcommand{\zz}{Z(3900)^{\pm}}
\newcommand{\pp}{\pi^{+}\pi^{-}}
\newcommand{\MM}{\mu^{+}\mu^{-}}
\newcommand{\jpsi}{J/\psi}
\newcommand{\ppjpsi}{\pi^{+}\pi^{-}J/\psi}
\def\pbnr{}
\def\speaker{Zhiqing Liu$^{1,2}$}
\def\onbehalfof{BESIII Collaboration and Belle Collaboration}
\def\title{Observation of $Z_c(3900)$ both by BESIII and Belle}
\def\affiliation{$^1$Johannes Gutenberg University of Mainz,\\
Johann-Joachim-Becher-Weg 45, D-55099 Mainz, Germany\\
$^2$Institute of High Energy Physics, \\
Chinese Academy of Sciences, Beijing, 100049}

\newcommand\pubnumber{\pbnr}
\newcommand\pubdate{\today}
\def\Title#1{\begin{center} {\Large #1 } \end{center}}
\def\Author#1{\begin{center}{ \sc #1} \end{center}}

\newcommand{\OnBehalf}[1]{\sbox0{#1}\ifdim\wd0=0pt
        {}
	\else
	{\\on behalf of #1}
	\fi}
\newcommand{\supportedBy}[1]{\sbox0{#1}\ifdim\wd0=0pt
        {}
	\else
	{\footnote{#1}}
	\fi}
\def\Address#1{\begin{center}{ \it #1} \end{center}}

\newcommand\pubblock{\includegraphics[width=5cm]{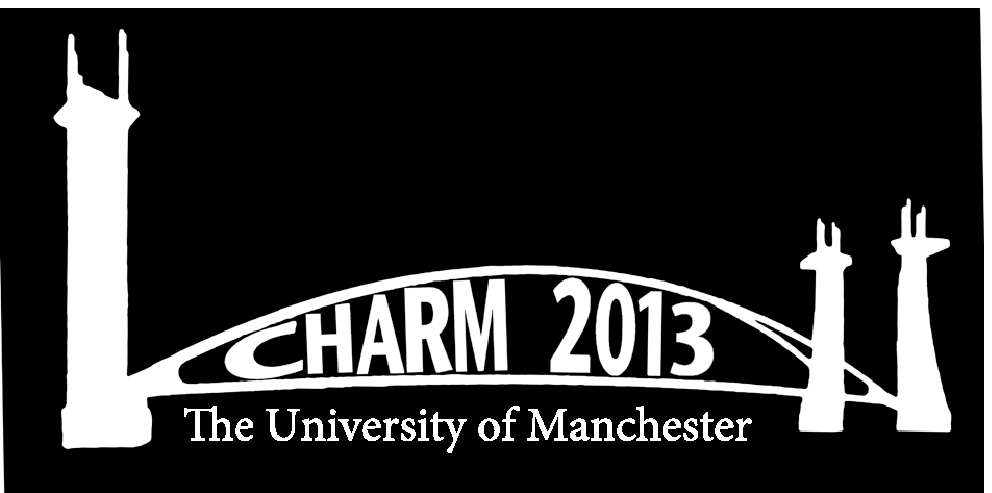}\hfill{\begin{tabular}{l} \pubnumber\\
         \pubdate  \end{tabular}}}
\newenvironment{Abstract}{\begin{quotation}  }{\end{quotation}}
\newenvironment{Presented}{\begin{quotation} \begin{center} 
             PRESENTED AT\end{center}\bigskip 
      \begin{center}\begin{large}}{\end{large}\end{center} \end{quotation}}
\def\Acknowledgements{\bigskip  \bigskip \begin{center} \begin{large}
             \bf ACKNOWLEDGEMENTS \end{large}\end{center}}
\def\venue{The 6$^{th}$ International Workshop on Charm Physics\\
(CHARM 2013)\\
Manchester, UK,  31 August -- 4 September, 2013}

\def\beq{\begin{equation}}
\def\eeq#1{\label{#1}\end{equation}}
\def\eeqn{\end{equation}}

\def\beqa{\begin{eqnarray}}
\def\eeqa#1{\label{#1}\end{eqnarray}}
\def\eeqan{\end{eqnarray}}

\let\bar=\overbar

\def\Dslash{\not{\hbox{\kern-4pt $D$}}}
\def\dslash{\not{\hbox{\kern-2pt $\del$}}}

\def\ee{e^+e^-}

\def\msb{{\bar{\ssstyle M \kern -1pt S}}}

\begin{document}
\begin{titlepage}
\pubblock

\vfill
\Title{\title}
\vfill
\Author{\speaker\supportedBy{}\OnBehalf{\onbehalfof}}
\Address{\affiliation}
\vfill
\begin{Abstract}
Using $525$~pb$^{-1}$ data collected with the BESIII detector at $e^+e^-$ central-of-mass energy 4.26~GeV, the BESIII Collaboration observed a charged charmoniumlike state $Z_c(3900)\to\pi^{\pm}J/\psi$ with mass $M[Z_c(3900)]=3899.0\pm 3.6 \pm 4.9$~MeV/c$^2$, and width $\Gamma[Z_c(3900)]=46\pm 10 \pm 20$~MeV. The significance of this state was estimated to be $>8\sigma$ in all kinds of systematic tests. Using $967$~fb$^{-1}$ data recorded by the Belle detector on or near $\Upsilon(nS)$, n=1,~2,~$\cdots$,~5 resonances, the $e^+e^- \to \pi^+\pi^-J/\psi$ process has been studied through initial-state-radiation (ISR) method. Except for the $Y(4260)$ resonance, an intermediate state
$Z(3900)^{\pm}$ has been observed in the $\pi^{\pm}J/\psi$ mass spectrum, with significance $>5.2\sigma$. The measured mass $M[Z(3900)^{\pm}]=3894.5\pm6.6\pm4.5$~MeV/c$^2$ and width $\Gamma[Z(3900)^{\pm}]=63\pm24\pm26$~MeV, shows good agreement with the BESIII measurement within errors, which means $Z_c(3900)$ and $Z(3900)^{\pm}$ is the same state.
\end{Abstract}
\vfill
\begin{Presented}
\venue
\end{Presented}
\vfill
\end{titlepage}
\def\thefootnote{\fnsymbol{footnote}}
\setcounter{footnote}{0}
%

\section{Introduction}
The well-known $\y$ resonance, was firstly observed by {\it BABAR} Collaboration in the ISR process $\ee\to \gamma_{ISR} \ppjpsi$~\cite{babary}, and confirmed by CLEO~\cite{cleoy} and Belle experiments~\cite{belley} subsequently. Being well above open charm threshold, the $\y$ resonance couples to lower lying charmonium state $\jpsi$ strongly, through its hadronic $\pp$ transition~\cite{xhm}. This is rather different from the already known conventional vector charmonium states in the same energy region, such as $\psi(4040)$, $\psi(4160)$ and $\psi(4415)$~\cite{pdg}, which mainly decay to open charm final states. The exotic features suggest the $\y$ might be not a conventional charmonium state~\cite{xyz-review}.

On the other hand, charged bottomoniumlike states $Z_b(10610)$ and $Z_b(10650)$ was observed in $\Upsilon(5S)\to\pp\Upsilon(1S,2S,3S)$ and $\Upsilon(5S)\to\pp h_b(1P,2P)$ decays~\cite{zb}. Similar structures were also reported in the charm sector, such as the $Z(4430)^{\pm}$ observed in the $\pi^\pm\psi(2S)$ mass spectrum of $B\to K\pi^\pm\psi(2S)$~\cite{z4430}, and two $Z^\pm$ states observed in the $\pi^\pm\chi_{c1}$ mass spectrum in $B\to K\pi^\pm\chi_{c1}$ decay~\cite{z1z2}. Inspired by these striking discoveries, we try to search for possible charged charmoniumlike states in $\y\to\ppjpsi$ decay.

The BESIII experiment has collected data at $\ee$ CM energy 4.26~GeV during December, 2012 to January 2013, which provides us an unique opportunity to study the $\y$ decay. Since its first $\y$ observation in 2007~\cite{belley}, the Belle experiment continued to accumulate data and the data sets are almost doubled since then. Thus, it also becomes necessary to analyze the $\y$ with full data sets. 

In this talk, we present the discovery of a new charged charmoniumlike state $\z$/$\zz$ through $\y\to\ppjpsi$ decay both at BESIII and Belle.

\section{Observation of $\z$ at BESIII~\cite{bes3-zc}}
The BESIII detector has collected 525~pb$^{-1}$ data at $\ee$ CM energy $(4.260\pm0.001)$~GeV. With this data sample, we analyze the $\ee\to\ppjpsi$ process. The Drift Chamber is used to catch 4 charged tracks ($\pp\ell^+\ell^-$), and the calorimeter is used to separate electrons and muons. Finally, we can effectively select $882\pm33$ signal events in $\jpsi\to\mu^+\mu^-$ mode, and $595\pm28$ signal events in $\jpsi\to\ee$ mode. The selection efficiency is measured by MC simulation, which gives $(53.8\pm0.3)$\% for $\MM$ mode, and $(38.4\pm0.3)$\% for $\ee$ mode, respectively. We use the published Belle~\cite{belley} and {\it BABAR}~\cite{babarnew} $\ee\to\ppjpsi$ cross section line shape to do radiative correction. And the radiative correction factor is determined to be $(1+\delta)=0.818$, which agrees well with QED prediction~\cite{rad}. Thus, the Born
order cross section for $\ee\to\ppjpsi$ at $\sqrt{s}=4.260$~GeV is measured to be $(64.4\pm2.4\pm3.8)$~pb in $\MM$ mode and $(60.7\pm2.9\pm4.2)$~pb in $\ee$ mode. The combined measurement gives $\sigma^{B}(\ee\to\ppjpsi)=(62.9\pm1.9\pm3.7)$~pb. The good agreement between BESIII, Belle~\cite{belley} and {\it BABAR}~\cite{babarnew} for $\ppjpsi$ cross section measurement confirms the BESIII analysis is valid and unbiased.

\begin{figure}[htb]
\centering
\includegraphics[height=1.4in]{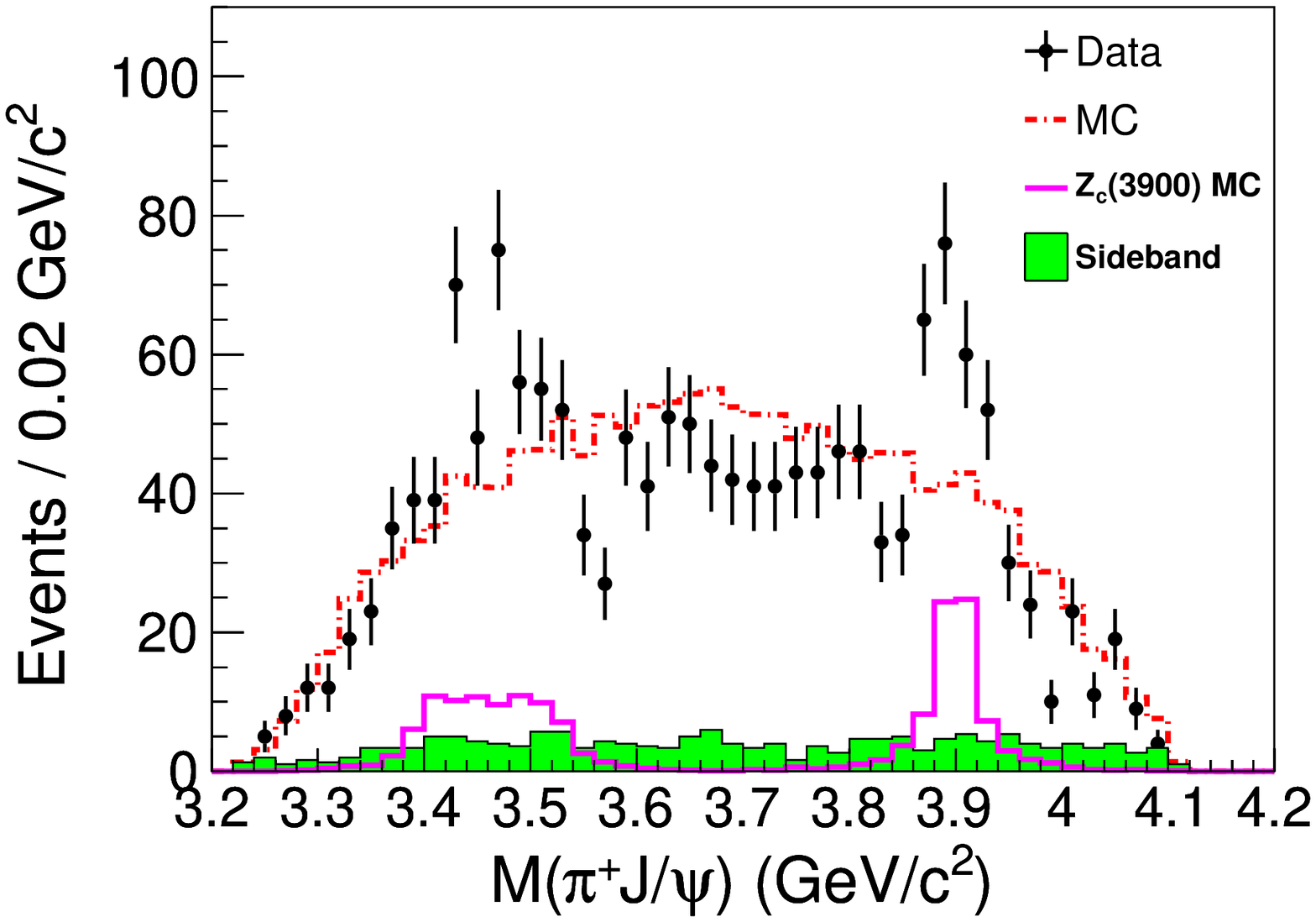}
\includegraphics[height=1.4in]{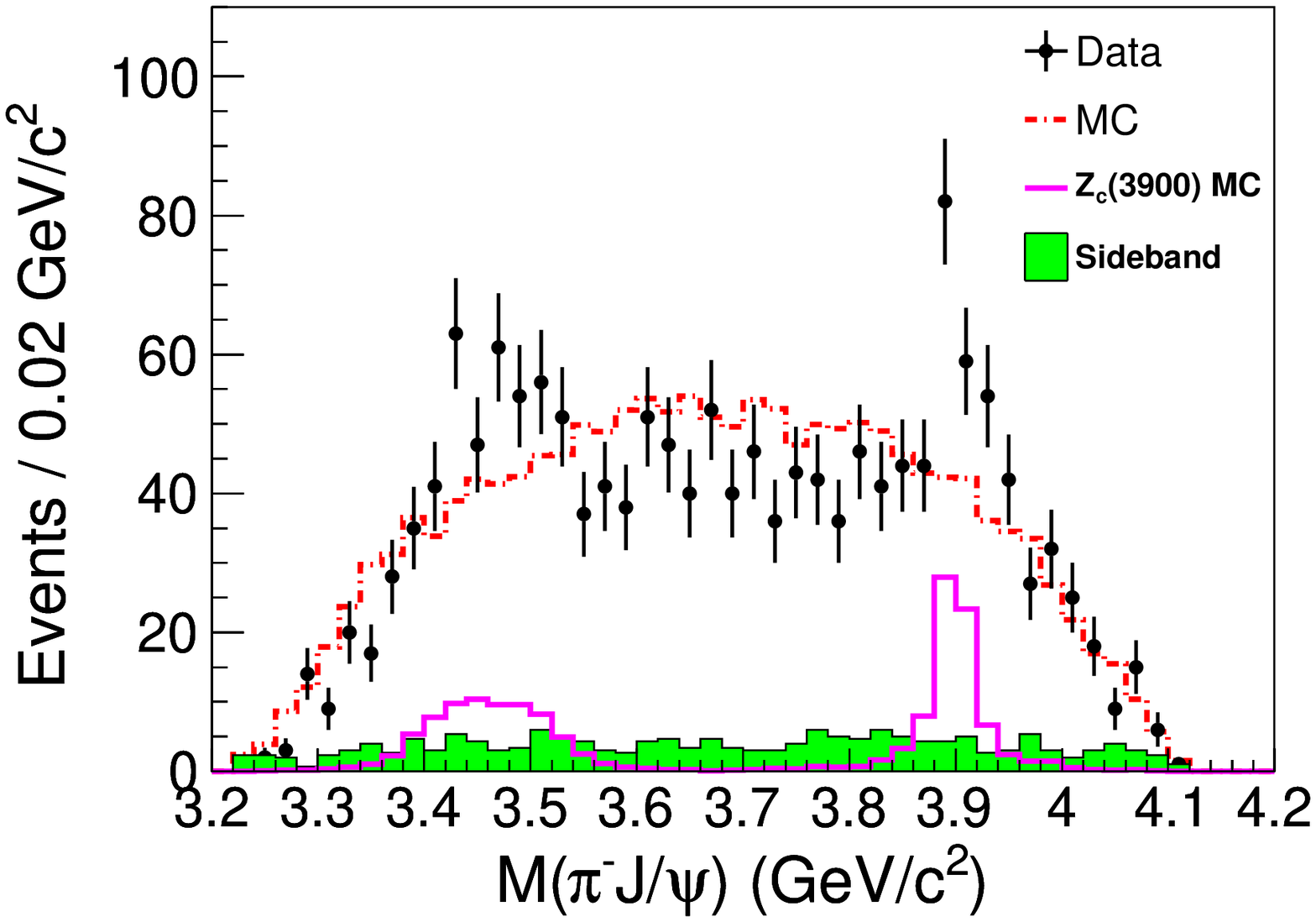}
\includegraphics[height=1.4in]{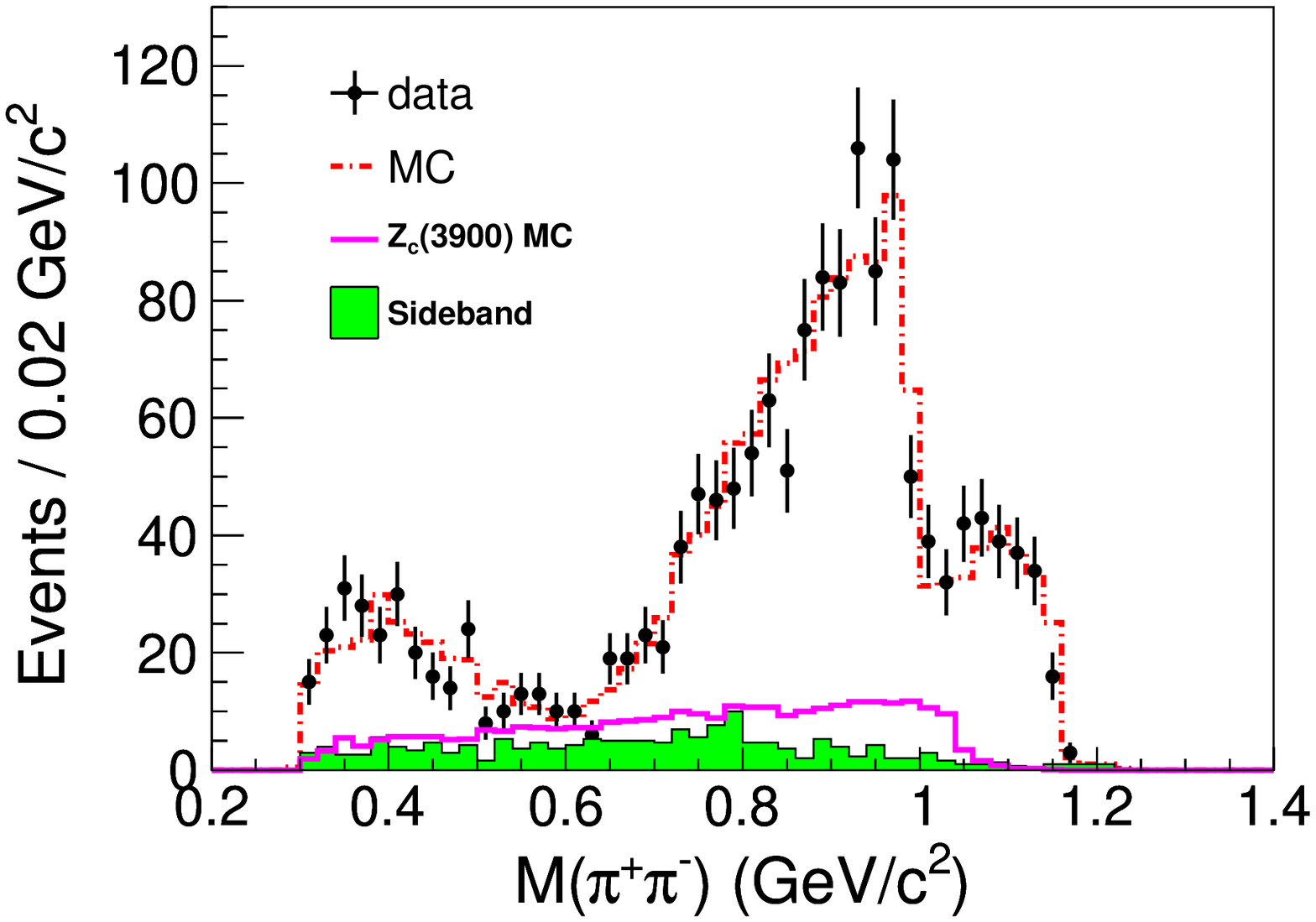}
\caption{$M(\pi^\pm\jpsi)$ and $M(\pp)$ invariant mass distributions for $\y\to\ppjpsi$ signal events at BESIII. Dots with error bars are data, green shaded histograms are non-$\ppjpsi$ background events estimated from $\jpsi$ mass sideband, red dotted-dashed histograms are MC simulation results from $\sigma(500)$, $f_0(980)$ and non resonant $S$-Wave $\pp$ amplitude, and the pink blank histograms shows a MC simulation of $\z$ resonance.
}
\label{bes-proj}
\end{figure}

After obtained the cross section, we turn to investigate the intermediate state in $\y\to\ppjpsi$ decays. The lepton pairs invariant mass is required to be between 3.08~GeV/c$^2$ and 3.12~GeV/c$^2$, in order to select $\jpsi$ candidates effectively. Finally we get 1595 $\ppjpsi$ signal events with a purity of $\sim$90\%. The Dalizt plot of $\y\to\ppjpsi$ signal events shows interesting structures both in the $\pp$ system and $\pi^\pm\jpsi$ system. Figure~\ref{bes-proj} shows the 1-dimensional projection of $M(\pi^\pm\jpsi)$ and $M(\pp)$ invariant mass distributions. In the $\pi^\pm\jpsi$ mass distribution, there is a clear peak at around 3.9~GeV/c$^2$ (called $\z$), while the wider peak in the lower mass side is a phase space reflection of $\z$. Such kind of effect can be simulated very well by MC events. For the $\pp$ mass distribution, there are also interested structures, which can be modeled well by $0^{++}$ resonance $\sigma(500)$, $f_0(980)$ and non-resonant $S$-wave $\pp$ amplitude. The $D$-wave $\pp$ amplitude is found to be small in data and they also do not form peaks in the $M(\pi^\pm\jpsi)$ mass spectrum.

\begin{figure}[htb]
\centering
\includegraphics[height=3in]{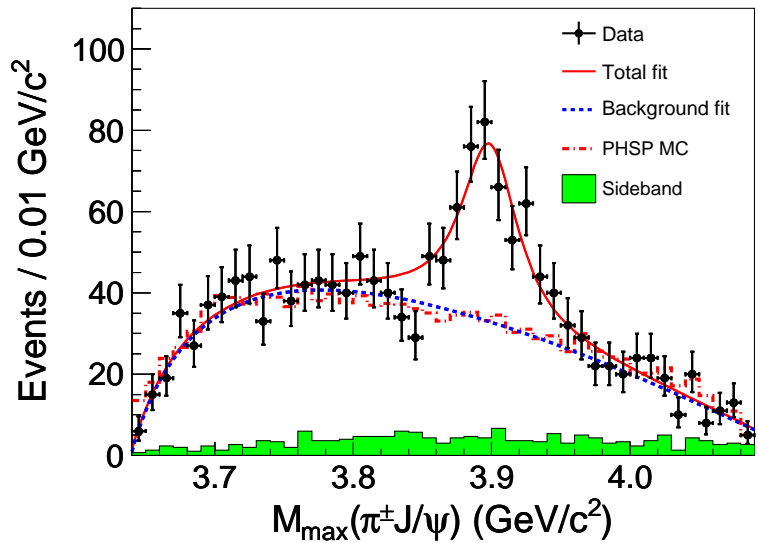}
\caption{Fit to the $M_{max}(\pi^\pm\jpsi)$ invariant mass distribution as described in the text. Dots with error bars are data, the red solid curve shows the total fit, dashed blue histogram is background contribution, and the red dotted-dashed histogram shows the distribution from pure phase space MC events. The green shaded histogram shows the non-$\ppjpsi$ background estimated from $\jpsi$ sideband.}
\label{bes-fit}
\end{figure}

To extract the resonant parameters of $\z$, we use 1-dimensional unbinned maximum likelihood fit to the $M_{max}(\pi^\pm\jpsi)$ mass distribution (the larger one of $M(\pi^+\jpsi)$ and $M(\pi^-\jpsi)$ mass combination in each event), which is an effective way to avoid $\z^+$ and $\z^-$ components cross counting. The signal shape is parametrized as an $S$-Wave Breit-Wigner function with free parameters, convolved with a Gaussian resolution function (parameters fixed according to the MC simulation value). The phase space factor $p\cdot q$ also has been considered, where $p$ is decay momentum of $\z$ in $\y$ CM frame and $q$ is the decay momentum of $\jpsi$ in $\z$ CM frame. The background shape is taken a shape as $a/(x-3.6)^b+c+dx$, where $a,b,c,d$ are free parameters and $x=M_{max}(\pi^\pm\jpsi)$. The efficiency curve also has been applied in the fit, and interference effect has not been considered currently. Figure~\ref{bes-fit} shows the fit results, with $M[\z]=(3899.0\pm3.6\pm4.9)$~MeV/c$^2$, and $\Gamma[\z]=(46\pm10\pm20)$~MeV. Here the first errors are statistical and the second systematic. The significance of $\z$ signal is estimated to be $>8\sigma$ in all kinds of systematic checks. The goodness of the fit is also found to be $\chi^2/ndf=32.6/37$. 

\section{Observation of $\zz$ at Belle~\cite{belle-zc}}

Using ISR method, the $\ee\to\ppjpsi$ cross section between 3.8~GeV and 5.5~GeV are measured, with 967~fb$^{-1}$ data collected by the Belle detector on or near $\Upsilon(nS)$, n=1,2,$\cdots$,5 resonances. The $\y$ resonant structure is observed with improved statistics, which agrees well with the previous Belle results~\cite{belley} and {\it BABAR} results~\cite{babary,babarnew}. We also use the ISR $\psi(2S)$ events to calibrate our analysis. Table~\ref{isr-psi} shows the measured ISR $\psi(2S)$ cross section in $\jpsi\to\ee$ and $\MM$ mode at different energy point, respectively. Good agreement is observed between our measurement and QED prediction~\cite{rad}, which also confirms our analysis is valid and unbiased.
\begin{table}[t]
\begin{center}
\caption{The measured ISR $\psi(2S)$ cross section (pb) in $\jpsi\to\ee$ and $\MM$ mode at different energy points, together with QED predictions~\cite{rad} with $\psi(2S)$ resonant parameters from PDG~\cite{pdg}.}
\begin{tabular}{cccc}
\hline\hline
$\sigma^{ISR}[\psi(2S)]$ & 10.02~GeV & 10.58~GeV & 10.87~GeV \\
\hline
$\ee$   & $16.75\pm0.85\pm1.01$ & $14.12\pm0.18\pm0.85$ & $13.79\pm0.44\pm0.83$ \\
$\MM$ & $16.63\pm0.54\pm0.87$ & $15.13\pm0.11\pm0.79$ & $13.33\pm0.25\pm0.70$ \\
QED    & $16.03\pm0.29$ & $14.25\pm0.26$ & $13.42\pm0.25$ \\
\hline\hline
\end{tabular}
\label{isr-psi}
\end{center}
\end{table}

\begin{figure}[htb]
\centering
\includegraphics[height=1.4in]{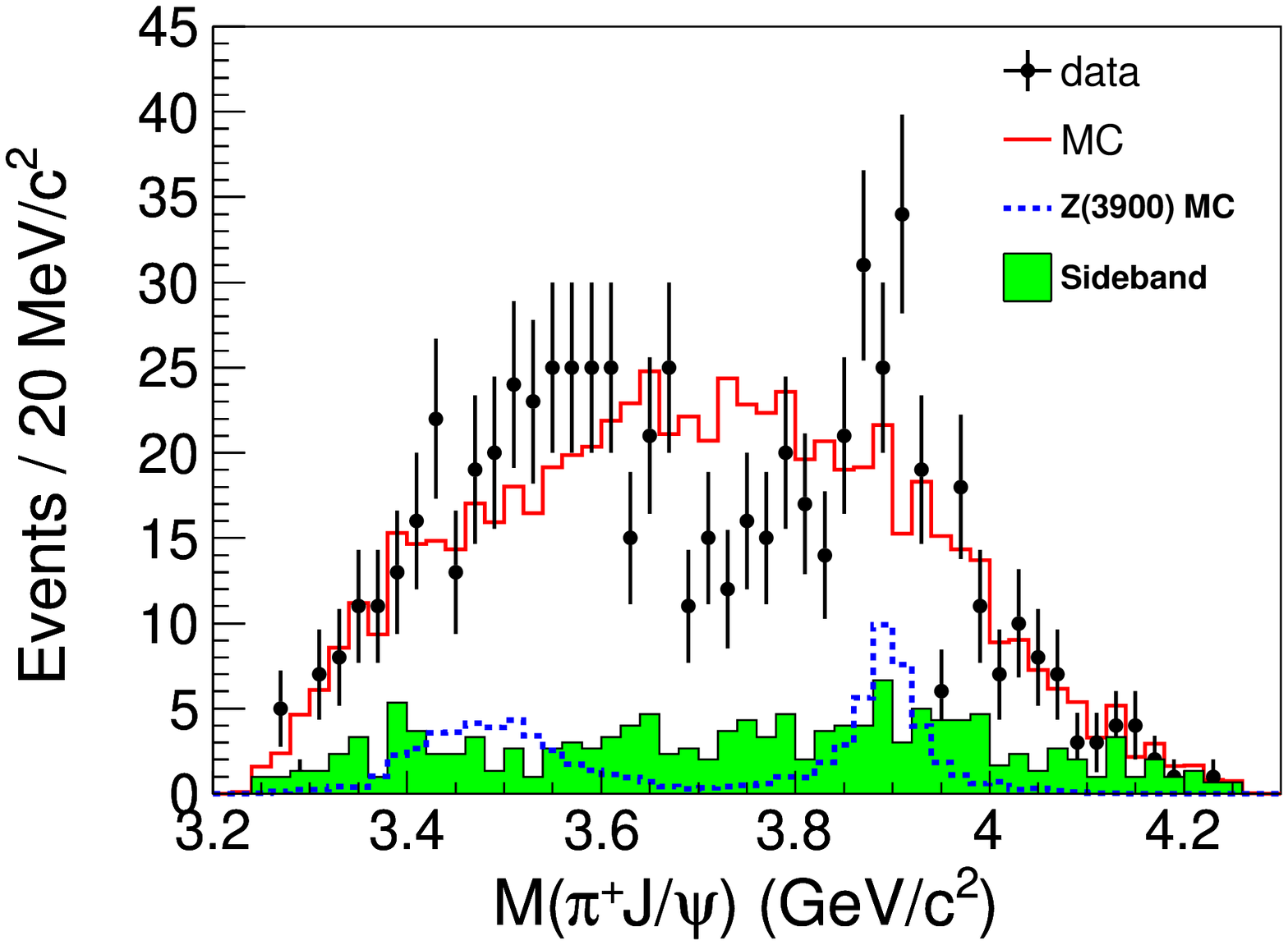}
\includegraphics[height=1.4in]{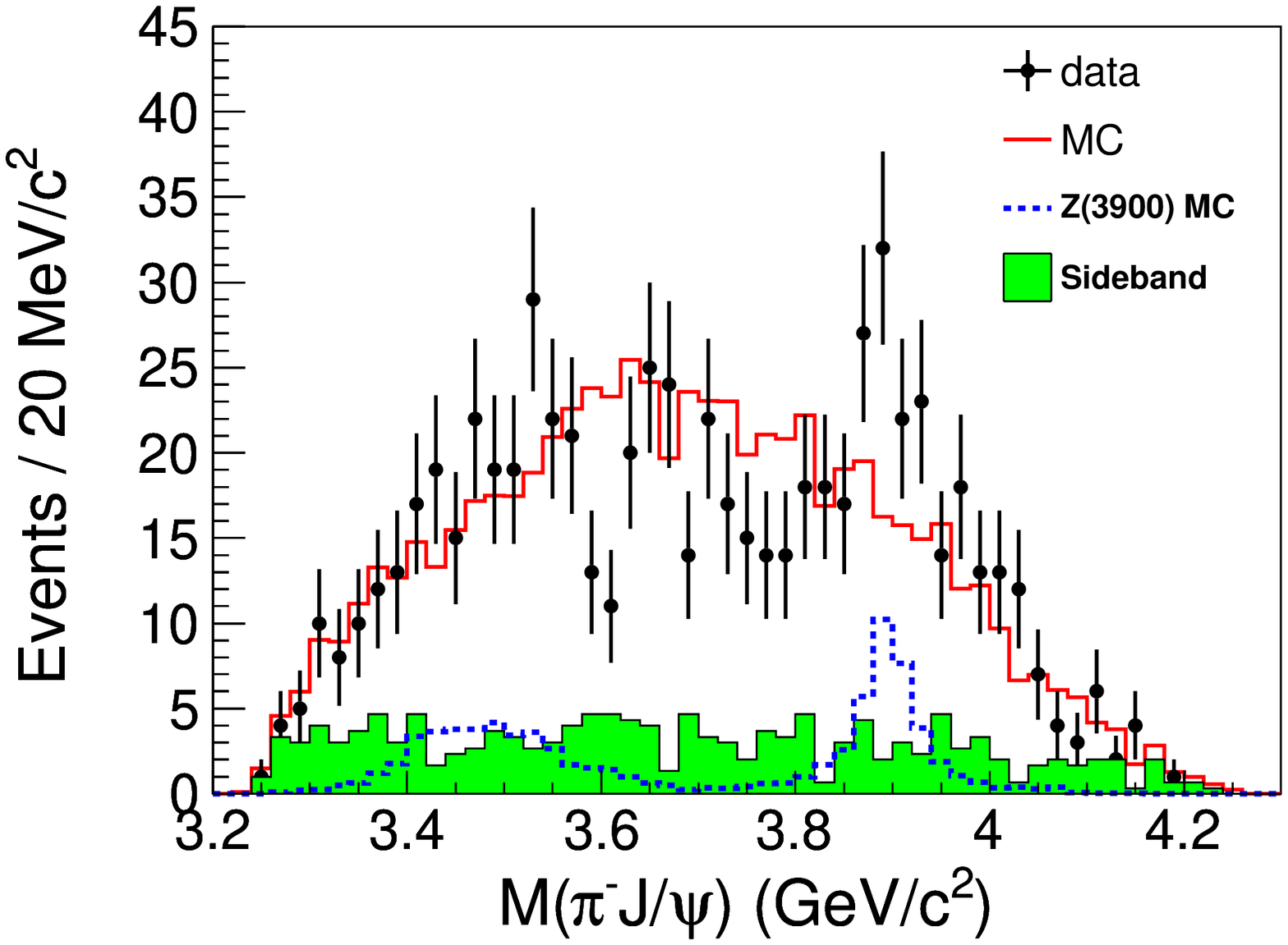}
\includegraphics[height=1.4in]{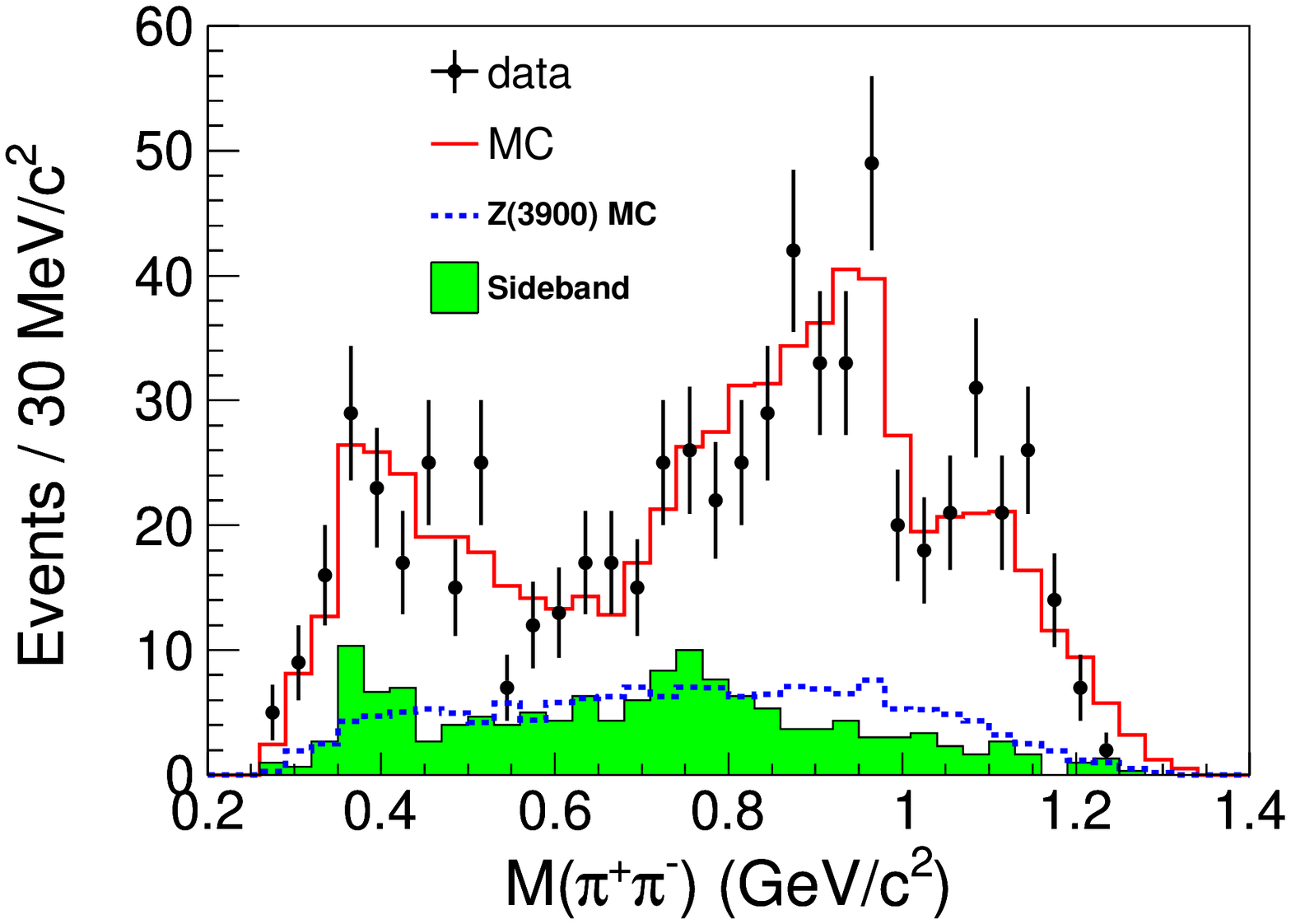}
\caption{The $M(\pi^\pm\jpsi)$ and $M(\pp)$ invariant mass distribution for $\y\to\ppjpsi$ signal events ($4.15<M(\ppjpsi)<4.45$~GeV/c$^2$). Dots with error bars are data, green shaded histograms are non-$\ppjpsi$ background events estimated from $\jpsi$ mass sideband, red solid histograms are MC distribution with parameters fixed to an $S$-wave $\pp$ amplitude fit, and the blue dashed histograms are MC simulation of $\zz$ signal events.}
\label{belle-proj}
\end{figure}

By requiring $4.15<M(\ppjpsi)<4.45$~GeV/c$^2$, we can select 689 $\y\to\ppjpsi$ events, with 139 background events which are estimated from $\jpsi$ mass sideband. Studying the Dalitz plot of this 3-body decay, we also find interesting structures both in the $M(\pi^\pm\jpsi)$ and $M(\pp)$ invariant mass distributions. Figure~\ref{belle-proj} shows the 1-dimensional projection of $M(\pi^\pm\jpsi)$ and $M(\pp)$, where a clear peak can be seen at $M(\ppjpsi)\sim$3.9~GeV/c$^2$ (called $\zz$). The relative wider peak at lower mass side are phase space reflection from $\zz$, which has been proved by MC simulation. In the $M(\pp)$ invariant mass distribution, there are clear $0^{++}$ resonance, such as $\sigma(500)$ and $f_0(980)$. We perform a partial wave analysis with $\sigma(500)$, $f_0(980)$, non-resonant $S$-wave $\pp$ amplitude, and $f_2(1270)$ amplitude, and find that $S$-wave (resonant and non-resonant) contributions is dominant in the $\pp$ system, which will not form peaks in the $M(\pi^\pm\jpsi)$ mass distribution. Figure~\ref{belle-proj} also shows the MC simulation results with amplitudes fixed to the partial wave fit results. We can see the $M(\pp)$ mass distribution is well described.

\begin{figure}[htb]
\centering
\includegraphics[height=3in]{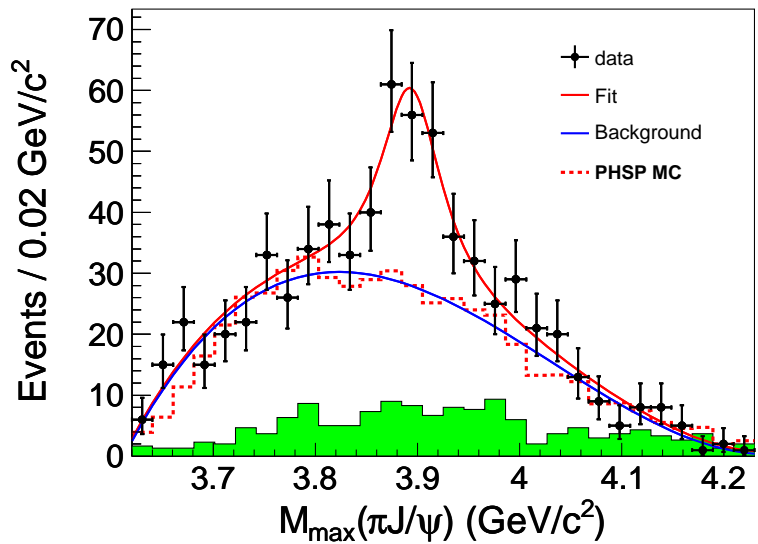}
\caption{Fit the $M_{max}(\pi^\pm\jpsi)$ invariant mass distribution with an $S$-wave Breit-Wigner function and cubic background term. Dots with error bars are data, green shaded histogram is non-$\ppjpsi$ background events estimated from $\jpsi$ mass sideband. The solid curves shows the fit results, and the dashed histogram shows the distribution from pure phase space $\ppjpsi$ events.}
\label{belle-fit}
\end{figure}

We try to extract the resonant parameters of $\zz$ using 1-dimensional unbinned maximum likelihood fit to the $M_{max}(\pi^\pm\jpsi)$ (the maximum of $M(\pi^+\jpsi)$ and $M(\pi^-\jpsi)$). Here interference effect is not considered. The signal shape is parametrized as $S$-wave Breit-Wigner function, convolving with a smearing Gaussian with parameters fixed according to MC simulation. The background is described by a cubic polynomial. The phase space factor has been included. The energy dependent detection efficiency also has been corrected. Figure~\ref{belle-fit} shows the fit results, which gives $M[\zz]=3894.5\pm6.6\pm4.5$~MeV/c$^2$ and $\Gamma[\zz]=63\pm24\pm26$~MeV, where the first errors are statistical and second systematic. The significance of $\zz$ signal is found to be larger than $5.2\sigma$ in all kinds of systematic checks.

\section{Summary and Discussion}

The $\z$/$\zz$ charged charmoniumlike state has been observed by the BESIII experiment at $\ee$ CM energy 4.26~GeV and the Belle experiment using ISR $\y\to\ppjpsi$ events independently. The measured mass and width by BESIII and Belle agree well with each other within errors, which indicates $\z$ and $\zz$ are the same state. It has been further confirmed by CLEO-c's data at $\ee$ CM energy 4.17~GeV~\cite{cleo-zc}. 
Table~\ref{zc-para} summarizes all the experimental results.
\begin{table}[t]
\begin{center}
\caption{$\z$/$\zz$ mass and width measurement from different experiments.}
\begin{tabular}{cccc}
\hline\hline
   & BESIII & Belle & CLEO-c's data \\ \hline
Mass~(MeV)  & $3899.0\pm3.6\pm4.9$ & $3894.5\pm6.6\pm4.5$ & $3886\pm4\pm2$ \\
Width~(MeV) & $46 \pm 10 \pm 20$          & $63 \pm 24 \pm 26$  &  $37 \pm 4 \pm8$ \\
\hline\hline
\end{tabular}
\label{zc-para}
\end{center}
\end{table}

$\z$ carries electric charged and couple to charmonium state. So it can not be a conventional charmonium state. The minimum quark content of $\z$ is a four quark state.
\Acknowledgements
I am grateful to the SFB1044 funding in Institute for Nuclear Physics, Johannes Gutenberg University of Mainz,
for supporting me the travel and the talk.

\end{document}